\documentclass[conference]{IEEEtran}
\IEEEoverridecommandlockouts

\usepackage{cite}
\usepackage{amsmath,amssymb,amsfonts, physics}
\usepackage{newtxtext,newtxmath}
\usepackage{algorithmic}
\usepackage{hyperref}
\usepackage{caption}
\usepackage{romannum}
\usepackage{float}
\usepackage{comment}
\usepackage{lipsum}
\usepackage{graphicx}
\usepackage{textcomp}
\usepackage{xcolor} 
\def\BibTeX{{\rm B\kern-.05em{\sc i\kern-.025em b}\kern-.08em
		T\kern-.1667em\lower.7ex\hbox{E}\kern-.125emX}}
\usepackage{fancyhdr}
\begin{document}
	
	\title{Nonlinearity Compensation and Link Margin Analysis of 112-Gbps Circular-Polarization Division Multiplexed Fiber Optic Communication System using a Digital Coherent Receiver over 800-km SSMF Link}

	\author{\IEEEauthorblockN{{A. K. M. Sharoar Jahan Choyon\textsuperscript{*,1},  Ruhin Chowdhury\textsuperscript{*}}}
		\IEEEauthorblockA{\textsuperscript{{*}}\textit{{Dept. of EECE, Military Institute of Science \& Technology, Dhaka, Bangladesh}}} 
				\textsuperscript{{1}}{choyonsharoar@gmail.com}\\
			
	}
	
	\maketitle

	\begin{abstract}
	Nonlinear effects have been considered as the major limitations in coherent optical (CO) fiber transmission system. DSP based CO receiver with digital backpropagation (DBP) method has recently facilitated the compensation of fiber nonlinear impairments as well as compensating dispersion of optical fiber. In this paper, a comprehensive design is presented for circular-polarization division multiplexed (CPDM) 8-quadrature amplitude modulation (8-QAM) with DSP based CO receiver for the fiber optic communication (FOC) system to investigate the impact of nonlinearities. We have examined the performance of nonlinearity compensating DSP based CO receiver with DBP method and demonstrated that it can be effectively employed to mitigate the intrachannel nonlinearities in CPDM 8-QAM FOC system over 800-km SSMF link. By effectively compensating the fiber nonlinearities, we analyze the performance of bit error rate (BER), optical signal to noise ratio (OSNR), optimum launch power, and investigate the link margin by evaluating OSNR margin for a specific launch power. Moreover, a CPDM technique helps as a very suitable means of maximizing the link capacity as well as enhancing the spectral-efficiency (SE) of the FOC system.

	\end{abstract}


\section{\textbf{Introduction}}

  Recently, the performance of FOC system is enhanced by introducing polarization division multiplexing (PDM) technique which divides the laser light into two orthogonal states of polarization (SOP), transmitting different signals over those polarization states ultimately doubling the SE of the system as well as the capacity \cite{1}-\cite{PDM-BD}. But as the demand increases, the channel capacity of optical communication must increase. To overcome this ever-growing demand and to tackle the challenge of increasing channel capacity, CPDM can be used. A CPDM system is basically an integration of two PDM systems involving right circular polarization (RCP) and left circular polarization (LCP), thus quadrupling the capacity and the SE of optical system \cite{rio}-\cite{CO-OFDM}.
  
The advancement of Erbium-doped fiber amplifier (EDFA), mitigation technologies of fiber nonlinearity, and dispersion management techniques has promptly enhanced the capacity of FOC system over the last few decades. Additionally, the link capacity and SE of optical fibers have also been improved by the research on higher-order modulation schemes. Because of lower bandwidth occupation for the identical bit rate with larger SE of the systems are typically more stable to polarization mode dispersion (PMD) and chromatic dispersion (CD), while in the higher bandwidth transmission systems, the fault tolerance to PMD and CD are mainly critical. Thus, to acquire a high SE, now the researcher has focused on coherent detection and higher-order modulation formats \cite{DSP}.

Due to the higher receiver sensitivity, coherent detection captured comprehensive research in the 1980s. Although the wavelength division multiplexing (WDM) and the invention of EDFA developed, research in CO communication discontinued at the end of 20\textsuperscript{th} century due to the complications of system employment, specifically, the complex implementation of the optical phase-locked loop. In practical realization of ordinary coherent receivers, the key impediments found to be the phase and polarization control. Luckily, in the electrical domain, phase and polarization control both can be implemented by DSP. Recently, CO communication has drawn considerable interest in the growing demand for link capacity and offered an exceptional and promising technique for implementing the high-capacity long-haul FOC systems \cite{4}-\cite{7}.

From the early 21\textsuperscript{st} century, using CO detection along with high bandwidth analog-to-digital converter (ADC), digital-to-analog converter (DAC), and DSP has augmented the feasible link capacity of FOC systems \cite{8}-\cite{11}.  In CO communication system, data is modulated in complex domain with the optical carrier signal and after experiencing several linear and nonlinear impairments, this modulated optical signal travels to the receiver end. Using CO detection, the complex field (phase and the magnitude) of received optical signal can be entirely obtained, and with the help of static and adaptive DSP, CD and PMD can also be completely compensated \cite{DSP}.

The opportunity of compensating the transmission impairments is one of the major advantages of using DSP with CO communication. Using DSP, CD and PMD compensations are now matured. While in CO communication systems, compensation of nonlinear impairments results in the essential capacity limit for the optical fiber transmission. Advanced DSP has also been developed in both frequency and time-domain equalization based adaptive digital filters. Moreover, in digital domain, DSP with dynamic equalizer can also perform polarization demultiplexing, tracking the rotations of SOP caused by the arbitrary changes in fiber birefringence \cite{DSP},\cite{nature}. The inter-symbol interference caused by PMD, and other static or linear distortions can be moderated by using the similar algorithm. The impact of CD can be compensated by using the static equalizer (time or frequency domain), while the impact of CD can be more strong to be mitigated in the dynamic equalizer, which provides advantages over optical dispersion compensating fibers (DCF) \cite{Rosa}.

	In this paper, we presents a complete design of 112 Gbps CPDM 8-QAM FOC system with nonlinearity compensating DSP based CO receiver to investigate the impact of nonlinearities and mitigation of nonlinear effects. By effectively compensating the fiber nonlinearities, we investigate the performance of BER, OSNR, optimum launch power, and link margin by evaluating OSNR margin over 800-km SSMF link. The organization of this paper is: Section \ref{sys} explicitly elucidates the CPDM 8-QAM CO-FOC system design with DSP based CO receiver, Section \ref{result} describes the results simulated by OptiSystem. Finally, this paper is concluded in Section \ref{conclusion} by compiling the analysis.

\section{\textbf{System Design \& DSP based CO Receiver}}\label{sys}
\subsection{\textbf{Design of CPDM 8-QAM CO-FOC System}}

	\begin{figure*}
	\centering
	\includegraphics[width=7in,height=9in,keepaspectratio]{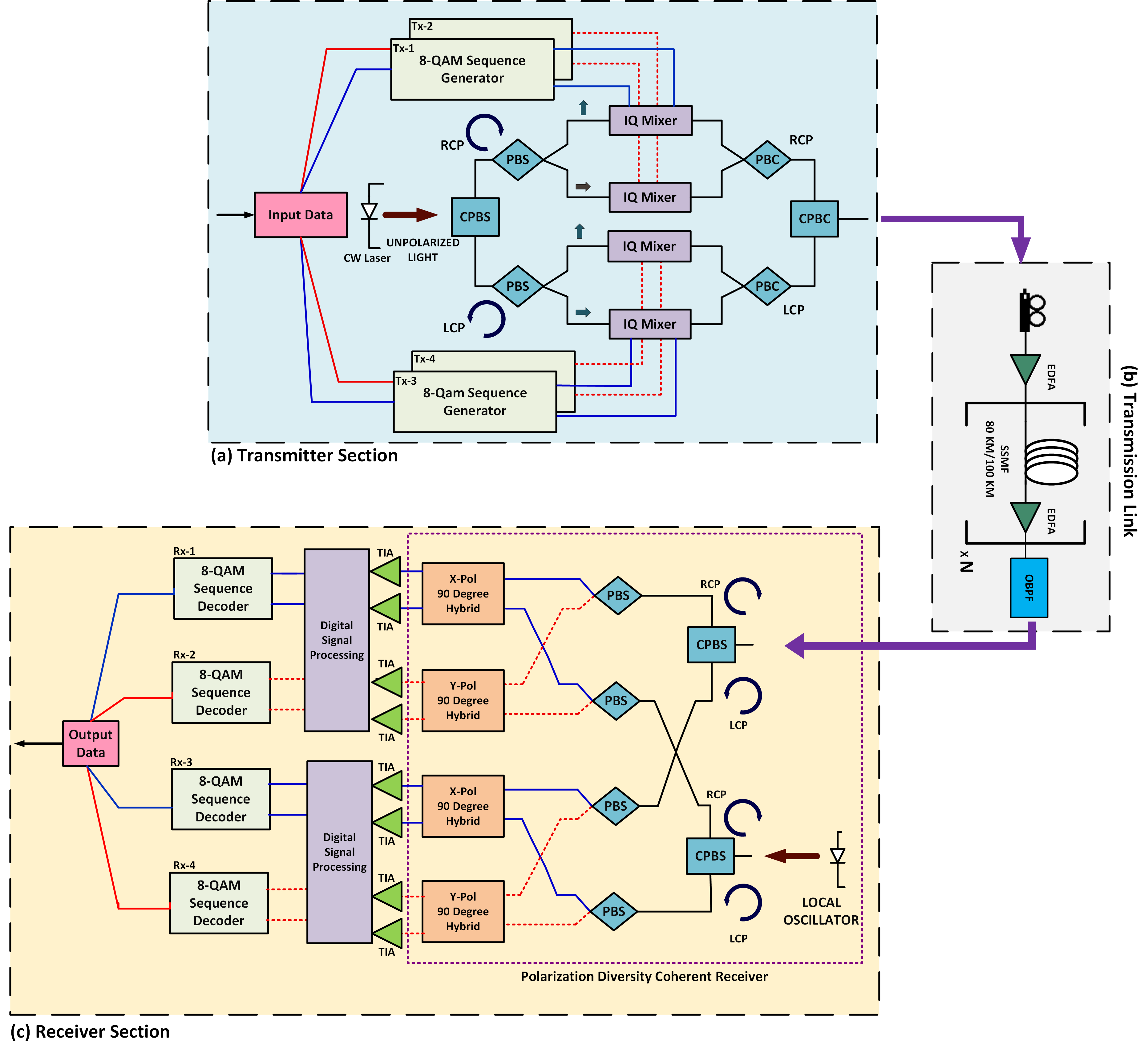} 
	\caption{\centering Design of CPDM 8-QAM CO-FOC System: (a) Transmitter section (b) Transmission link (c) Receiver section with DSP. }\label{design}	
    \end{figure*} 
    
	\begin{figure*}
	\centering
	\includegraphics[width=7in,height=3.5in,keepaspectratio]{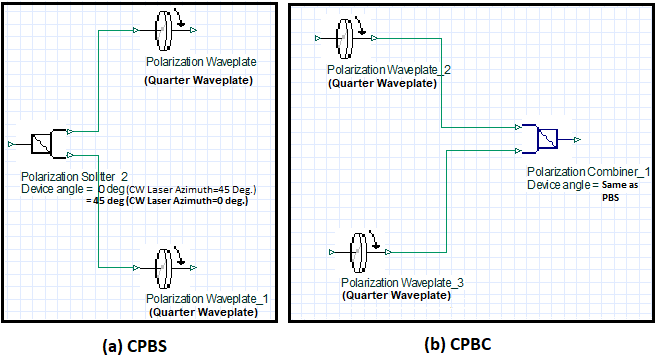} 
	\caption{\centering Design of (a) CPBS (b) CPBC using Quarter Waveplate. }\label{CPBS-C}	
    \end{figure*}

	\begin{figure*}
	\centering
	\includegraphics[width=7in,height=7in,keepaspectratio]{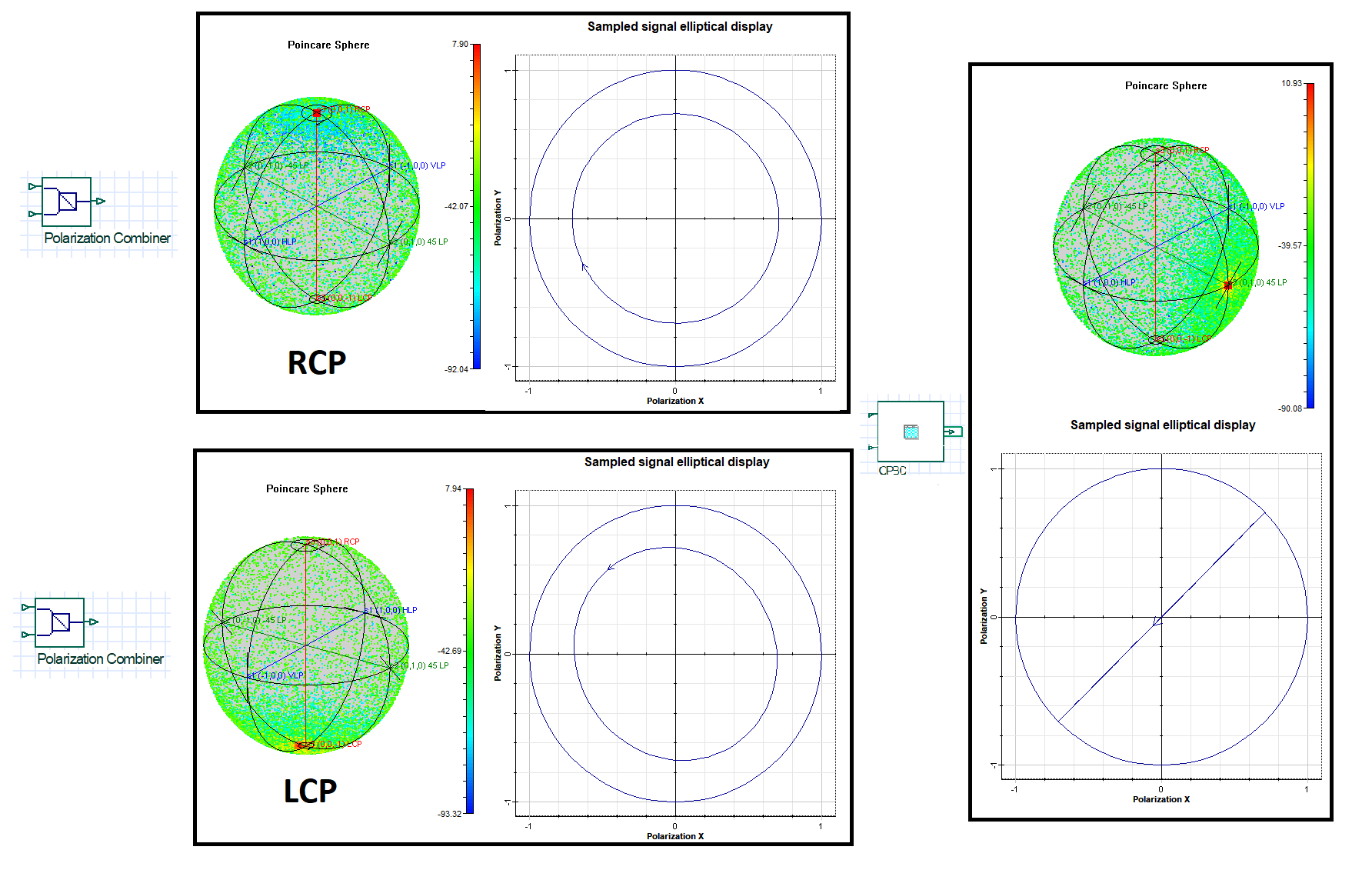} 
	\caption{\centering SOP of modulated optical signal after IQ mixer. }\label{poincare}	
    \end{figure*}

	\begin{figure}
	\centering
	\includegraphics[width=3.5in]{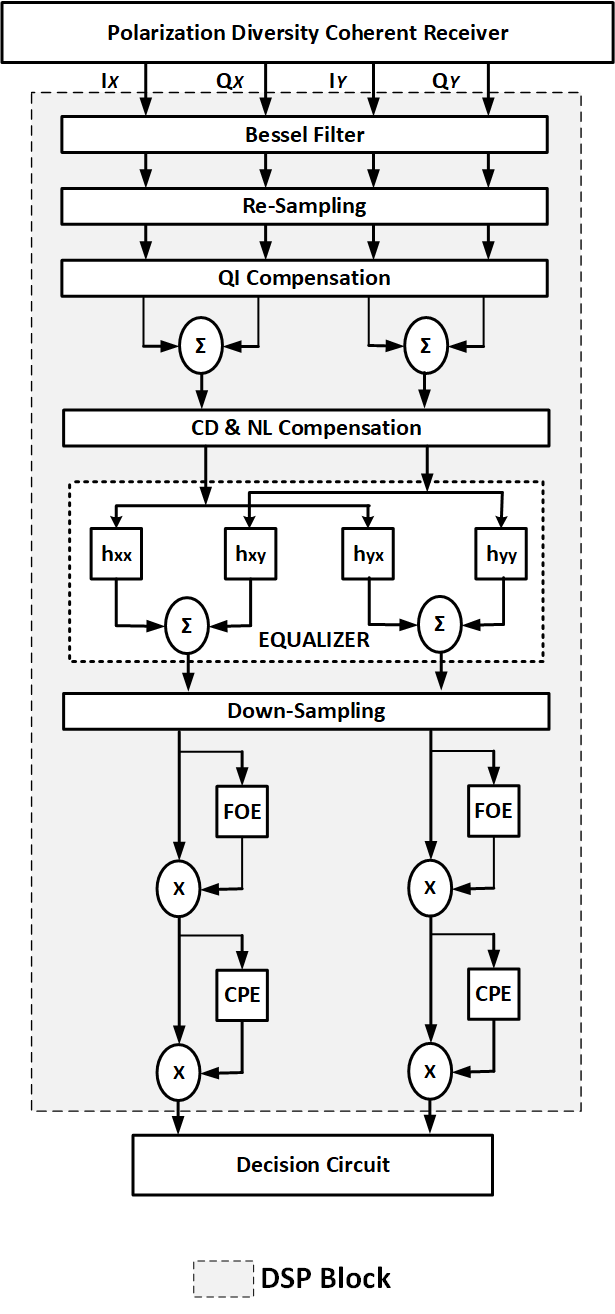} 
	\caption{\centering Schematic Diagram of DSP Block. }\label{DSP block}	
    \end{figure} 
    
	\begin{figure*}
	\centering
	\includegraphics[width=6.5in,height=2in,keepaspectratio]{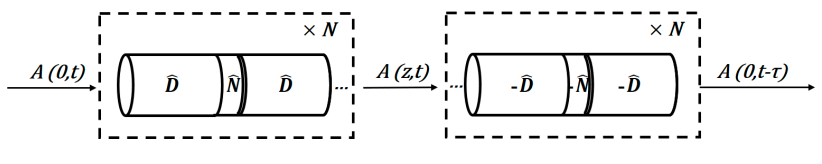} 
	\caption{\centering Schematic of DBP using the symmetric split-step method (SSM). }\label{DBP}	
    \end{figure*} 

A comprehensive system design for CPDM 8-QAM CO-FOC system is depicted in Fig. \ref{design}. A source ( CW Laser diode ), a circular polarization beam splitter (CPBS), a circular polarization beam combiner (CPBC), two PBS, two polarization beam combiner (PBC) and four optical IQ mixers form the transmitter. The CW Laser power is taken as 20 dBm, operating wavelength is set to 1550 nm and is operated in 45 degrees azimuth. The input laser power  is divided into two circularly polarization states-Right Circular and Left Circular- using a CPBS. see Fig. \ref{CPBS-C}. At the transmitter section, each of the output of a CPBS is fed into a PBS. The PBSs divide the circularly polarized (RCP \& LCP) laser power into two orthogonal polarization states (Horizontal \& Linear). These power signals are fed into optical IQ Mixers used as carriers. Thus, two independent PDM systems consisting of 4 independent channels create a CPDM system \cite{rio}-\cite{CO-OFDM}.

Meanwhile, the 112 Gbps information signal is divided into four equal segments each transmitting 28 Gbps 8-QAM modulated data are fed into four optical IQ mixers \cite{IQ mix}. Therefore, each PBS generates two modulated data (X \& Y Polarized). In this model, using two PBS, we get two sets of X and Y Polarized modulated data. Each PBC combines a set of X and Y Polarized modulated data. The two PBCs produce RCP and LCP light respectively and a CPBC combines these into linear polarized light which is sent through the fiber optic channel. The transmission fiber line is an 800 km re-circulating loop consisting of ten 80 km/eight 100 km SSMF with an Erbium-doped optical fiber amplifiers (EDFA). An optical band-pass filter (OBPF) is used to restrain ASE noise and is added as a post amplifier with 100 GHz bandwidth. Poincare sphere explains the SOP of modulated optical signal which is generated by using OptiSystem, see Fig. \ref{poincare}.

As a result, 28 Gbps signal is applied to each QAM generator and the achieved signal rate of the CPDM transmitter becomes 112 Gbps using four IQ mixers whereas, the signal rate of a PDM transmitter would have been 56 Gbps using only two IQ mixers. The SE and the capacity of a CPDM transmitter is thus doubled comparing to a PDM transmitter without altering the bandwidth of the transmitter as explained.

 A polarization diversity coherent optical receiver receives the signal. It is then recovered by splitting into RCP and LCP components by a CPBS, see Fig. \ref{design}(c). The separated components are then further split into orthogonal components. Similarly, a combination of a CPBS and two PBS divides the LO (CW Laser source with same parameters used during transmission) signal into its orthogonal components and later mixed with the received signal. Using balanced detectors and transimpedance amplifiers (TIA), the optical signals are converted into electrical signals, amplified and changed to voltage respectively. Analog to digital converter (ADC) converts the analog electrical signals to digital at DSP block and finally decoded by an 8-QAM decoder to retrieve the original data. Detailed system design parameters are elucidated in TABLE \ref{para}. 

\subsection{\textbf{DSP based CO Receiver}}

The DSP block, depicted in Fig. \ref{DSP block}, after coherent detection in OptiSystem provides various advantages and recovers the incoming signals. Each DSP block used dual-polarization X and Y channels. The preprocessing stage of each DSP block includes three functions: adding noise to signal, enabling DC block, and normalizing the received signal to 8-QAM grid. The following stage is the signal recovery stage, which consists of eight functions described below:

\textit{\textbf{(\romannum{1}) Bessel Filter:}} Using a 4th order Bessel filter, band noise is removed with a bandwidth of 3 dB and 28 GHz.

\textit{\textbf{(\romannum{2}) Re-sampling:}} This function re-samples the sampled input signal at 2 samples/symbol rate with cubic interpolation to modify the input signal.

\textit{\textbf{(\romannum{3}) QI Compensation:}} When the signal passes through the transmission line it experiences distortions due to misalignment of the CPBS and PBS. Moreover, setting improper modulator bias voltage, imbalanced photodiode responsivity, and flaws in the optical 90-degree hybrid create phase and amplitude imbalances within the in-phase (I) and quadrature (Q) signals \cite{QI}. QI compensation helps to restore the signal by recovering the balance of phase and amplitude.

\textit{\textbf{(\romannum{4}) CD Compensation:}} Similar to the QI compensation function, the signal travelling through the optical fiber is compensated for CD by using the CD Compensation function. CD is polarization independent and a static phenomenon. A dispersion compensating filter is used to remove CD which can be either in frequency or time domain \cite{CD}. 

\textit{\textbf{(\romannum{5}) Nonlinearity (NL) Compensation:}} While propagating through optical fiber, optical signal suffers from nonlinear and linear impairments. This limits the transmission capacity. Therefore, eliminate these impairments and restore original signal, researchers have provided various solutions.  Currently, researchers have successfully solved linear impairments (CD and PMD) effectively, so the challenge lies with nonlinear impairments which is the limitation of FOC systems’ capacity. To reach the required OSNR along with improving SE and increasing the transmission distance high-order modulation format and high signal power are used. But increasing the signal power increases nonlinear impairments which can be reduced in the receiver section by using DSP in the digital domain. Therefore, mitigating these impairments is possible and in turns it increases the channel capacity. Besides, nonlinear impairments i.e. cross-phase modulation (XPM) and four-wave mixing (FWM) can also be reduced so long the adjacent channels are received.

Currently, digital backward propagation (DBP) is the most favorable approach to compensate nonlinearities. The nonlinear Schrödinger equation (NLSE) which gives a solution in the digital domain can compensate for the deterministic impairments completely in fiber optic transmission. The DBP method is illustrated in Fig. \ref{DBP}. The distorted signal in the actual transmission link goes through the virtual fiber to remove the distortions and retrieve the original signal. A digital model with opposite propagation parameters detects the received signal and processes it. Backward propagated NLSE can be expressed as \cite{DBP}:

\begin{equation}
    \frac{\partial A}{\partial (-z)} = (\hat{D} + \hat{N})A
\end{equation}
\begin{equation}
    \hat{D} = -j\frac{\beta_{2}}{2}\frac{{\partial}^2}{\partial t^2} + \frac{\beta_{3}}{6}\frac{{\partial}^3}{\partial t^3} -\frac{\alpha}{2}
\end{equation}
 
\begin{equation}
    \hat{N} = j\gamma {\abs{A}}^2
\end{equation}

Where, A = complex electric field of the received signal, $\hat{N}$ = nonlinear operators, $\hat{D}$ = linear operators, $\alpha$ = attenuation, $\beta_2$ = first-order group velocity dispersion, $\beta_3$ = second-order group velocity dispersion, and $\gamma$ = fiber nonlinear coefficients.

Selecting NLSE propagation parameters is one of the crucial factors to achieve favorable performance of DBP algorithm. A common solution of NLSE is the split-step Fourier method (SSFM). Both linear and nonlinear impairments can be reduced by exploiting backward propagated NLSE equation as pre and post-compensation. The use of DBP as a post-compensation along with coherent receivers is most common method used for diminishing nonlinearities. One of the advantages of using DBP is that it can be implemented to any modulation format since it operates on the complex envelope of the electrical field. But DBP cannot compensate stochastic noise source, for instance, ASE noise which limits the performance of this method. This method is only suitable for compensating deterministic signal distortions.

\textit{\textbf{(\romannum{6}) Timing Recovery:}} Determining the accurate symbol sampling time as well as synchronizing the symbols are the two key roles of this algorithm.

\textit{\textbf{(\romannum{7}) Adaptive Equalizer (AE):}} Even after CD compensation, some CD and PMD still remains in the signal. These residual distortions as well as inter-symbol interference are compensated by using AE. In case of dual-polarization system, polarization demultiplexing is achieved using the butterfly structure. This structure of the equalizer helps to equalize the linear impairments along with demultiplexing the PDM signals with significant crosstalk. Some of the proposed polarization demultiplexing algorithms are radius directed equalization (RDE), constant modulus algorithm (CMA), recursive least square (RLS), least mean square (LMS), etc \cite{AE1}-\cite{AE2} in which, two-stage CMA-RD algorithm is used.

\textit{\textbf{(\romannum{8}) Frequency Offset Estimation (FOE) and Carrier Phase Estimation (CPE):}} Usually, the phase and frequency of the LO wave must match the phase and frequency of the signal carrier as the primary requirement of CO detection. However, in reality, the transmitter and the LO laser frequencies are not in perfect consistent with each other. This happens due to some environmental variation and fabrication imperfection of optical devices. Moreover, phase noise is introduced by the linewidth of the lasers which is a Wiener process and it harms the phase-modulated signal greatly. The main objective of this algorithm in CO detection is to mitigate the carrier frequency offset and phase noise impairments by processing a discrete data sample sequence \cite{DSP}. This function is followed by the blind phase search (BPS) algorithm \cite{BPS} which recovers and removes the residual phase mismatch between the LO and the main signal.

\section{\textbf{Results \& Discussions}} \label{result}

We investigate the system performance as well as the mitigation of fiber nonlinearities on 112-Gbps CPDM 8-QAM 800Km fiber optic transmission system in terms of BER, OSNR, OSNR margin (OSNR\textsubscript{margin}), required OSNR (OSNR\textsubscript{req}) for specific BER, optical launch power for 80Km/100Km per span SSMF length. We use OptiSystem 17.0 and MATLAB to design and simulate the system. The parameters used for the simulation are depicted in TABLE \ref{para}.

	\begin{table}
	\centering
	\caption{Design parameters and their values.}\label{para}
\begin{tabular}{c c}
			\hline
			\textbf{Parameter} & \textbf{Value} \\ \hline
			Bit Rate & 112 Gb/s\\
			CW Laser Power & 20 dBm\\
			CW Laser linewidth & 0.1 MHz\\
			Propagation Length & 800 Km\\
			Per span SSMF length & 80 km \& 100 km\\
			Number of loops & 10 (80 km) \& 8 (100 km)\\
			Optical carrier wavelength & 1550 nm\\
			CW LO laser power & 20 dBm\\
			CW  LO Laser linewidth & 0.1 MHz\\
			
			Modulation type & 8-QAM\\
			Fiber nonlinear index of refraction & 26$\times$10\textsuperscript{-21} m\textsuperscript{2}/W  \\
			Effective area of fiber & 80 $\mu$m\textsuperscript{2}\\
			
			Optical filter BW & 100 GHz\\
			Dark current & 10 nA\\
			Thermal Bandwidth & 10 GHz\\
			PIN Photodiode Responsivity & 0.95 A/W\\
			Temperature	& 298 K \\
		    TIA Gain & 20 dB\\
			Load Resistance &	50 $\Omega$ \\
			Spontaneous emission factor & 1.4\\
			EDFA Gain & 20 dB\\ 
			Attenuation co-efficient & 0.2 dB/Km\\
			Dispersion co-efficient  & 16.75 ps/nm.km \\
			Residual dispersion slope & 0.075 ps/nm\textsuperscript{2}.km \\
		    Nonlinear ration & 0.48 \\
		    Nonlinear compensation & 0.76 \\
		    EDFA noise figure & 4 dB \\

			\hline 
		\end{tabular}
	
\end{table}

	\begin{figure*}
	\centering 
	\includegraphics[width=7in,height=3.5in,keepaspectratio]{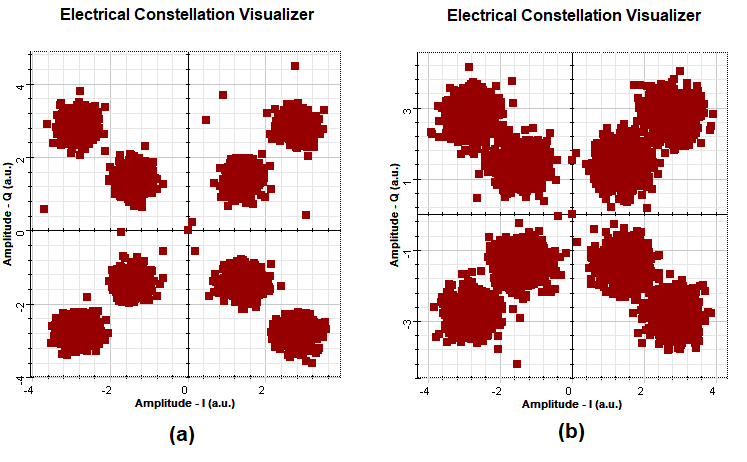} 
	\caption{\centering Received signal constellation diagram after 560-km link distance: (a) without nonlinearities (b) with nonlinearities.}\label{const}
\end{figure*}

\begin{figure*}
	\centering 
	\includegraphics[width=7in,height=3.5in,keepaspectratio]{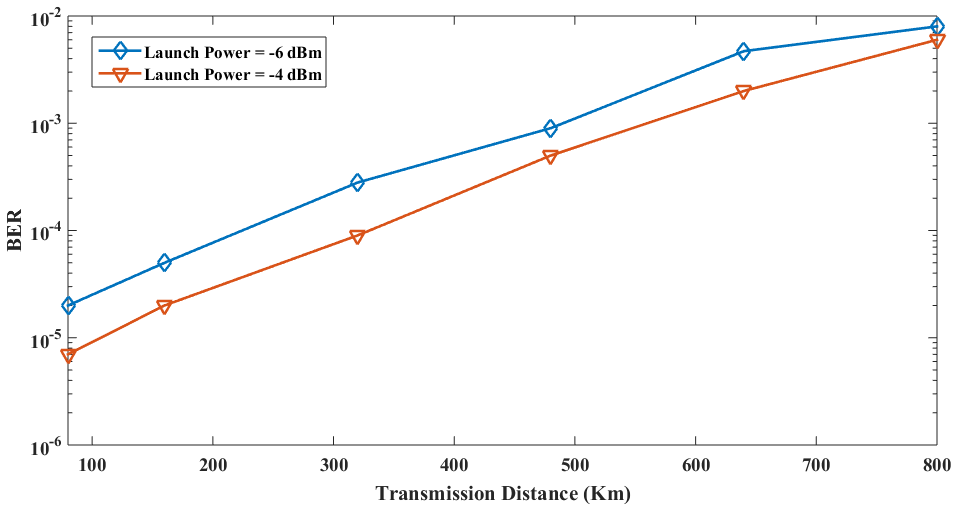} 
	\caption{\centering BER vs. Transmission distance.}	\label{BERvL}
\end{figure*} 

\begin{figure*}
	\centering 
	\includegraphics[width=7in,height=3.5in,keepaspectratio]{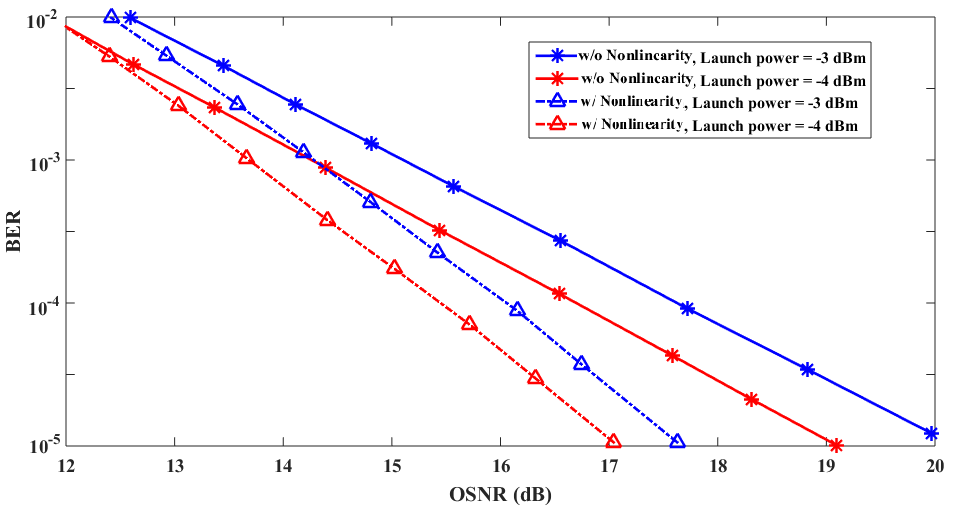} 
	\caption{\centering BER vs. OSNR.}	\label{BERvOSNR}
\end{figure*}

\begin{figure*}
	\centering 
	\includegraphics[width=7in,height=3.5in,keepaspectratio]{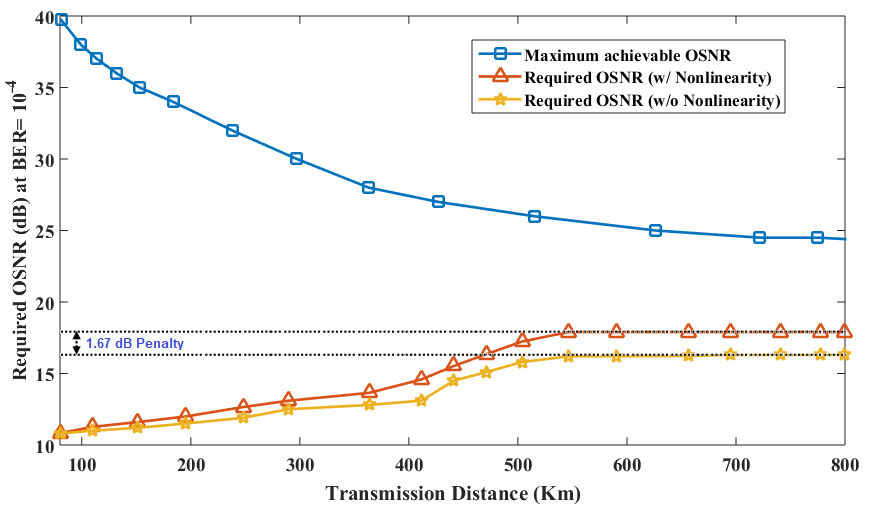} 
	\caption{\centering OSNR\textsubscript{req} vs. Transmission distance.} 	\label{reqOSNR}
\end{figure*}

\begin{figure*}
	\centering 
	\includegraphics[width=7in,height=3.5in,keepaspectratio]{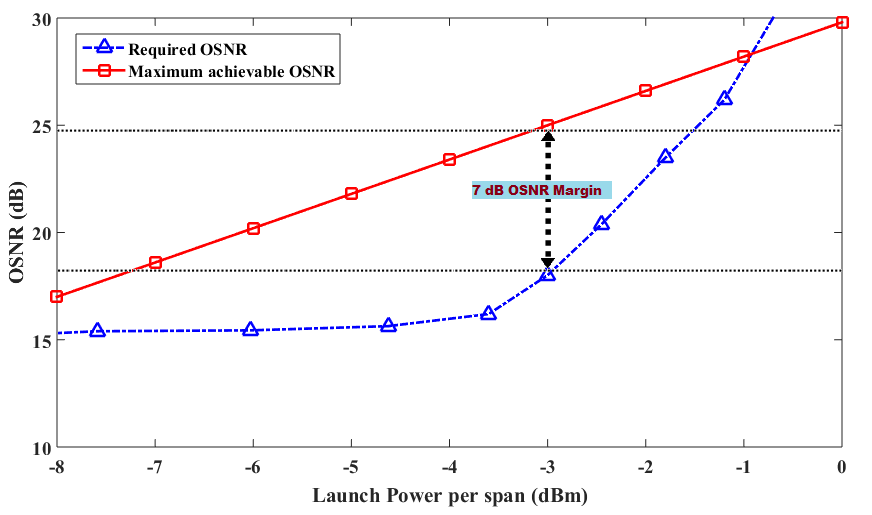} 
	\caption{\centering OSNR vs. Launch power for 80-km per span SSMF length.}  \label{span80}
\end{figure*}

\begin{figure*}
	\centering 
	\includegraphics[width=7in,height=3.5in,keepaspectratio]{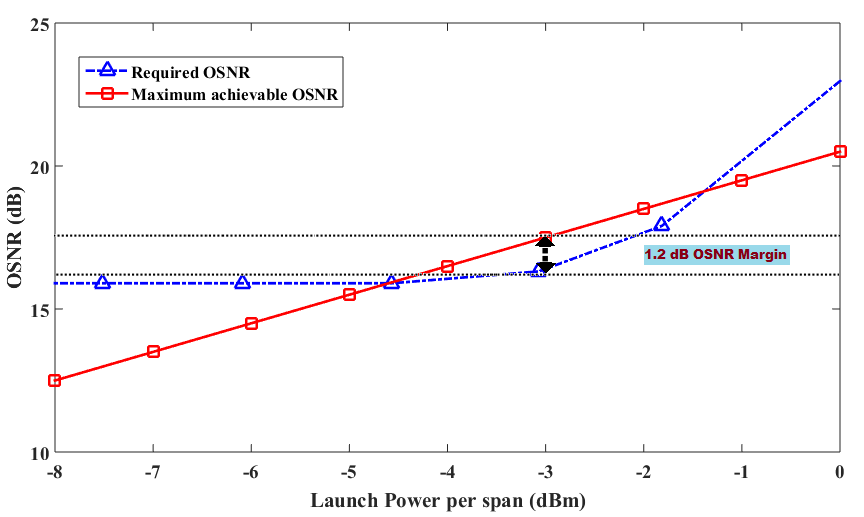} 
	\caption{\centering OSNR vs. Launch power for 100-km per span SSMF length.}  \label{span100}
\end{figure*}

\begin{figure}
	\centering 
	\includegraphics[width=3.5in]{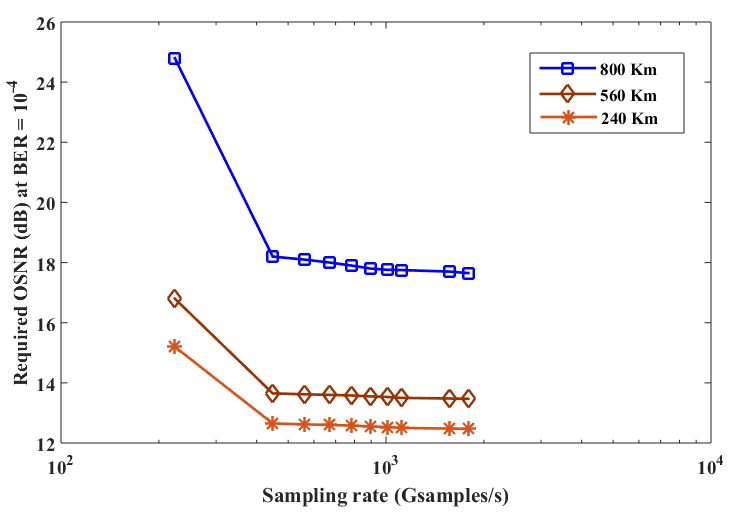} 
	\caption{\centering OSNR\textsubscript{req} vs. Sampling rate.}  \label{samp}
\end{figure}

	Fig. \ref{const} shows the received signal constellation diagram of 112-Gbps CPDM 8-QAM FOC system after 560Km transmission length with and without considering the fiber nonlinearities. Here we consider 80Km SSMF length for per span transmission. As the fiber nonlinearities are added to the system, signals are scattered from its reference level. The impact of optical launch power on the performance of BER as a function of transmission length is depicted in Fig. \ref{BERvL}. As the transmission length increases, BER also deteriorates and an improvement of signal launch power from -6dBm to -4 dBm enhances the BER performance. Fig. \ref{BERvOSNR} explicates the impact of the presence of fiber nonlinearity on the performance of BER and OSNR for different signal launch power of -3 dBm and -4 dBm. OSNR degrades approximately 2.7 dB due to the presence of fiber nonlinearity to accomplish a certain BER of $10^{-5}$ at -3 dBm optical launch power considering per span 80Km SSMF length.

The performance of OSNR in absence of fiber nonlinearity, the OSNR\textsubscript{req} is not affected by the compositions of fiber link. But, if we consider fiber nonlinearity, it would be necessary for the fiber optic transmission system to outline the optical signal launch power as well as per span fiber length to support for adequate OSNR\textsubscript{margin}. Fig. \ref{reqOSNR} depicts the effects of fiber nonlinearity for the span of 80km SSMF at -3dBm per span signal launch power as a function of transmission link distance and the OSNR\textsubscript{req} (with and without the nonlinearity of fiber) with the maximum achievable OSNR. It is conferred that an additional penalty of 1.67dB is required for the system lengths less than 800Km resulted from incorporating the fiber nonlinearity in transmission simulation, see in Fig. \ref{reqOSNR}. Furthermore, the maximum achievable OSNR deteriorates due to the rise in fiber length. Regarding an optical laser signal of 1550nm wavelength, 0.1MHz linewidth, and 4dB noise figure, the maximum achievable OSNR was around 26dB at system transmission length of 500Km and 24dB at 800km (80km SSMF span) respectively; see in Fig. \ref{reqOSNR}. From this graph, it is noted that, at 800km transmission length, the maximum achievable OSNR was higher than the required value by 7dB.

Fig. \ref{span80} illustrates the required OSNR with the maximum achievable OSNR as a function of optical launch power at 800Km transmission length considering 80Km per span SSMF length keeping other system parameters identical as in Fig. \ref{reqOSNR}. From the graph it explicates that due to the effects of nonlinearity, the OSNR\textsubscript{req} of 15.2dB presented a negligible penalty for an optical signal launch power below -5dBm. On the other hand, the penalty influenced by the nonlinearity of fiber raised significantly when optical launch power was larger than -3dBm, although the maximum achievable OSNR augmented uniformly with the optical launch power. Consequently, for an optimal launch power of around -3dBm, the OSNR\textsubscript{margin} reaches nearly about 7dB. It is noteworthy that, a supplementary self-phase modulated compensator \cite{nature} might be applied to moderate the nonlinearities, in which the OSNR\textsubscript{margin} of the systems may further be improved. However, it may need further complicated electronic devices.

Similarly, Fig. \ref{span100} depicts the OSNR\textsubscript{req} with the maximum achievable OSNR as a function of optical launch power at 800Km transmission distance considering 100Km per span SSMF length. It is noteworthy that the suitable performance was unfeasible at specific optical launch powers. In such instances, noise produced by in-line amplification was ignored while simulations for yielding the OSNR\textsubscript{req} to be stabilized. It is conferred from the graph that, in contrast to the configuration of 80Km per span structure at that equal launch power, the OSNR\textsubscript{req} applying 100Km per span was reduced due to the decreased effects of nonlinearity from the shortened number of spans for higher launch powers. Nonetheless, the maximum achievable OSNR reduced greatly, and hence, the OSNR\textsubscript{margin} was reduced to approximately 1.2dB.

The simulations using the OptiSystem above assume 448 Gsamples/s as a sampling rate. For evaluating the effect of sampling rate considering fiber nonlinearities, Fig. \ref{samp} explicates the relationship of OSNR\textsubscript{req} for a certain BER of $10^{-4}$ and sampling rate for different transmission distances, up to 800Km. The graph presents the adverse impact of decreasing the sampling rate as well as demonstrates that it is flexible for shorter system length. For instance, at a 240Km transmission distance, the OSNR\textsubscript{req} at the lowest sampling rate of 224 Gsamples/s (sample per bit 2) in comparison to a 1120 Gsamples/s (sample per bit 10) reference raised by 2.7dB, although at 800Km, the OSNR\textsubscript{req} raised by almost 7dB at that similar condition.

\section{\textbf{Conclusion}}\label{conclusion}
	A comprehensive design is presented for the FOC system to incorporate CPDM 8-QAM with DSP aided CO receiver over 800-km SSMF link to investigate the impacts of nonlinearities as well as compensating the fiber nonlinearities. It is noteworthy from this investigation that OSNR degrades approximately 2.7 dB due to the presence of fiber nonlinearity to acquire a BER of 10\textsuperscript{-5} at -3 dBm optimum optical launch power considering per span 80-km SSMF length. By effectively compensating nonlinear impairments and selecting an optimal signal launch power of -3 dBm, the OSNR margin reaches nearly about 7dB for 800-km transmission using 80-km SSMF span length. Moreover, a design of CPDM scheme helps to maximize the link capacity and enhance the SE of the FOC system.

\section{\textbf{Acknowledgement}}
This research did not get any special grant from funding authorities in the public, commercial, or not-for-profit sectors.

\end{document}